\documentclass[fleqn,twoside]{article}
\usepackage{espcrc2}

\usepackage{graphicx}

\newcommand{\be}{\begin{equation}}
\newcommand{\ee}{\end{equation}}
\newcommand{\bea}{\begin{eqnarray}}
\newcommand{\eea}{\end{eqnarray}}

\newcommand{\AmS}{{\protect\the\textfont2
  A\kern-.1667em\lower.5ex\hbox{M}\kern-.125emS}}

\title{Cosmological constraints on Neutrino - Dark Matter interactions}

\author{G. Mangano\address{INFN, Sezione di Napoli,
Dipartimento di Scienze Fisiche, Universit\`a di Napoli {\it
Federico II}}}

\begin{document}

\begin{abstract}
I summarize the results of a recent analysis where the
cosmological effects of interactions of neutrinos with cold Dark
Matter (DM) is investigated. This interaction produces
diffusion-damped oscillations in the matter power spectrum,
analogous to the acoustic oscillations in the baryon-photon fluid.
I discuss the bounds from the Sloan Digital Sky Survey on the
corresponding opacity defined as the ratio of neutrino-DM
scattering cross section over DM mass, and compare with the
constraint from observation of neutrinos from supernova 1987A.
\vspace{1pc}
\end{abstract}

\maketitle

\section{INTRODUCTION}

The most favored candidates for dark matter (DM) are cold,
collisionless massive particles, which are non-relativistic for
most of the history of the Universe and so can cluster
gravitationally during matter domination. Candidates for these
dark matter particles can be found in supersymmetric extensions of
the standard electroweak model. However, observations on galactic
and sub-galactic scales may conflict with the predictions from
numerical simulations and analytic calculations. Indeed, cold and
collisionless dark matter models seem to predict an excess of
small-scale structures \cite{silk1}, and numerical simulations
predict far more satellite galaxies in the Milky Way halo than are
observed.

Several solutions have been proposed to explain these
discrepancies. In this paper, I summarize some results for
scenarios where light (MeV) dark matter interacts with standard
species such as leptons \cite{boehm2}. In particular, if DM and
neutrinos interact, so that there was an epoch in the very early
universe during which they were strongly coupled we expect to see
some effect on the Large Scale Structure (LSS) power spectrum
\cite{prd}. In fact, DM perturbations that entered the horizon
during this period would then be erased because of diffusion
damping, and the suppression scale will depend on the dark
matter--neutrino interaction. Even if only a fraction of the dark
matter interacts with neutrinos, a pattern of oscillations in the
matter power spectrum arises, much like the oscillations in the
baryon-photon fluid.

\section{BOUNDS ON NEUTRINO DARK MATTER INTERACTION}

We consider here the case of a non-self-conjugated scalar particle
$\psi$ with mass $m_{DM}$ in the MeV range as a candidate for DM
\cite{boehm2} and an interaction term dictated by the following
lagrangian density
\begin{equation} {\cal L}_{int}= g \overline{F}_R \nu_L \psi + h.c.
\label{lscalar} \end{equation} where $F$ is a spinor field. Other
possible couplings, for example via the exchange of an
intermediate vector-boson field $U_\mu$, are extensively discussed
in \cite{prd}.

In the range of neutrino temperature $T \leq MeV$ we are
interested in (earlier evolution affects in fact, the power
spectrum at very small scales with wavenumber larger than 10$^5$ h
Mpc$^{-1}$) the opacity , i.e. the thermally averaged
$\psi$-neutrino scattering cross section is \be \frac{\langle
\sigma_{DM-\nu} |v| \rangle }{m_{DM}} \sim \frac{|g|^4}{m_{DM}}
\frac{T^2}{(m_F^2-m_{DM}^2)^2} \equiv Q_2 \frac{1}{a^2}
\label{sigmafs} \ee with $a$ the scale factor normalized to unity
at the present time. If the $\psi$ and $F$ fields are degenerate
in mass the low-energy transfer scattering cross section has
instead a Thomson behavior, \be \frac{\langle \sigma_{DM-\nu} |v|
\rangle }{m_{DM}} \sim \frac{|g|^4}{m_{DM}^3} \equiv Q_0
 \label{sigmafsdeg} \ee

The effect of neutrino-DM scattering processes is to modify the
standard Euler equations for the velocity perturbations, which now
keep an additional term proportional to the opacity.

\begin{figure}[htb]
\includegraphics[height=12pc]{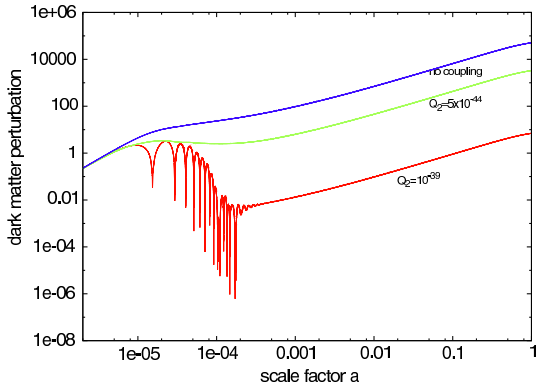}
\caption{Dark matter perturbations of $k=1.04\,h\,$Mpc$^{-1}$. The
opacity $Q_{2}$ is in unit of cm$^{2}$~MeV$^{-1}$ \cite{prd}. }
\label{fig:1}
\end{figure}

Consider for example the case $\langle \sigma_{DM-\nu}|v| \rangle
\propto a^{-2}$ (similar considerations can be made for the
constant cross section of Eq. (\ref{sigmafsdeg})). In Fig.
\ref{fig:1}, I show what happens when a perturbation of wavenumber
$k=1.04\,h$ Mpc$^{-1}$ enters the horizon for different values of
$Q_2$. If the coupling is zero the mode enters the horizon in the
radiation-dominated era, and it starts to grow first
logarithmically and then linearly with the expansion factor
(during matter domination). When the same mode enters the horizon
with $Q_{2}=5\times\,10^{-44}$ cm$^{2}$~MeV$^{-1}$, the growth is
nearly zero during the radiation epoch, while the mode starts
growing linearly with the scale factor during matter domination,
since the coupling with neutrinos becomes negligible at this stage
for this value of $Q_2$. For a larger $Q_{2}=10^{-39}$
cm$^{2}$~MeV$^{-1}$, when the perturbation enters the horizon,
dark matter is coupled with neutrinos and this results in a series
of oscillations until decoupling is reached. Notice that the
amplitude of oscillations decreases near decoupling due to
diffusion damping for the dark matter--neutrino fluid.
\begin{figure}[htb]
\includegraphics[height=12pc]{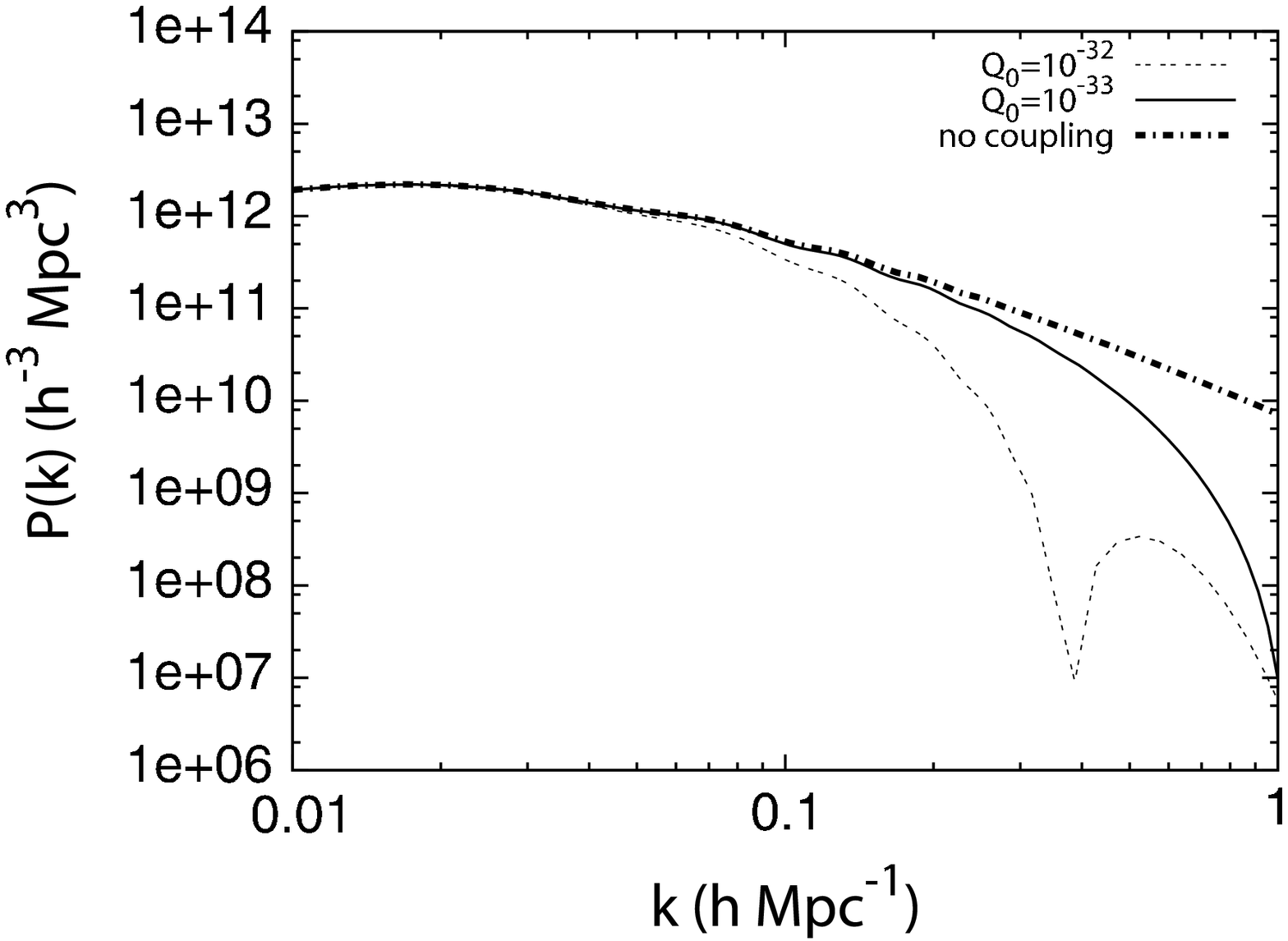}
\includegraphics[height=12pc]{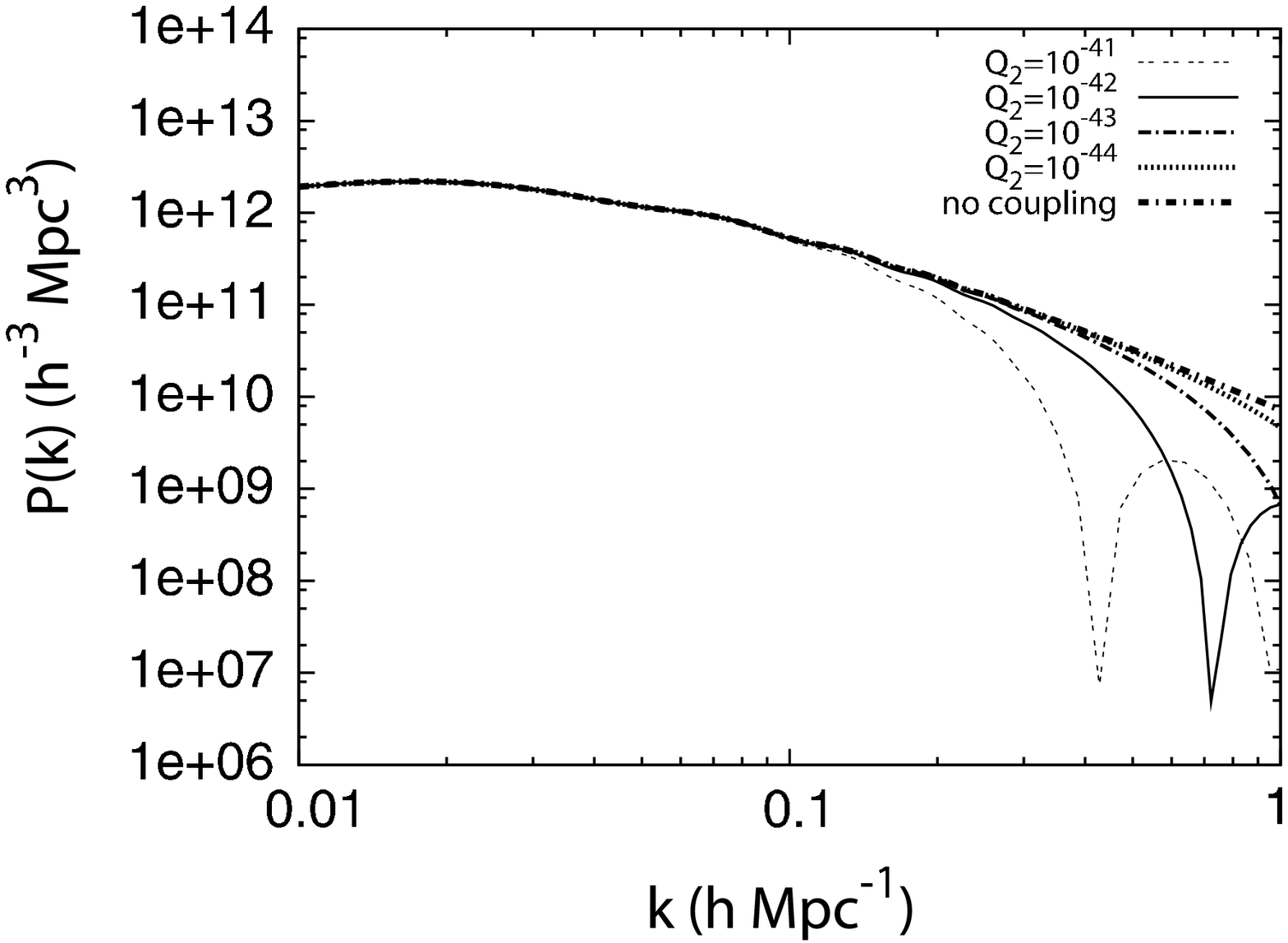}
\caption{Matter power spectra for different opacities $Q_{0}$ (top
panel) and $Q_2$ (bottom panel) \cite{prd}.} \label{fig:2}
\end{figure}
In Fig. \ref{fig:2}, I show several matter power spectra for
different values of the DM-neutrino interaction for both the case
considered. The effect can be seen on small scales. Larger
couplings will correspond to later epochs of neutrino-DM
decoupling and to a damped oscillating regime on larger scales. On
the other hand, even considering values of $Q_2$ or $Q_0$ which
are already at odds with current clustering data, there is only a
small enhancement in the small-scale CMB anisotropies which are
therefore, very weakly affected by the neutrino-DM interactions
\cite{prd}.

In order to bound the strength of DM coupling to neutrinos, one
can use the real-space power spectrum of galaxies in the Sloan
Digital Sky Survey (SDSS) using the data and window functions of
the analysis of Ref.~\cite{thx}. To compute the likelihood
function for the SDSS, the analysis has been conservatively
restricted in \cite{prd} to a range of scales over which the
fluctuations are assumed to be in the linear regime ($k < 0.2
h^{-1}\rm Mpc$) marginalizing over a bias $b$ considered to be an
additional free input. Assuming a cosmological concordance model
with $\Omega_{\Lambda}=0.70$ and $\Omega_{DM}=0.25$ which produces
a good fit to current CMB data, one obtains \bea Q_{2} &\le&
10^{-42}\,\mbox{cm}^2~\mbox{MeV}^{-1} \\ Q_{0} &\le& \,10^{-34}
\,\mbox{cm}^{2}~\mbox{MeV}^{-1} \eea at the $2 \sigma$ confidence
level. Comparing these results with Eq.~(\ref{sigmafs}), we see
that for couplings $g$ of order one this bound is saturated if
both $m_{DM}$ and $m_F$ are of the order of MeV. Smaller values of
$g$ imply lighter masses for the DM and the intermediate exchanged
particle in the scattering process. In view of the bound on
neutrino coupled dark matter from Big Bang Nucleosynthesis found
in \cite{serpico}, $m_{DM} \geq 1$ MeV, these values are already
disfavored, so the LSS constraint above is not further
constraining the model. More stringent bounds might be obtained by
studying the LSS power spectrum at larger wavenumbers, $k \geq 0.2
\,h$ Mpc$^{-1}$, taking into account the nonlinear behavior of
perturbations for very small scales. On the other hand, the result
for $Q_0$ which we recall corresponds to an intermediate particle
and DM mass degeneracy, is more severely constraining $m_{DM}$. We
get in this case $m_{DM} \geq 10 g^{4/3}$ GeV.

Notice that if one assumes that DM couples to neutrinos strongly
enough to produce observable effects that can be constrained by
CMB and LSS observations, one have to abandon the idea that relic
DM density formed via the usual mechanism based on freezing of DM
annihilation processes at temperatures $T \sim m_{DM}$, for
annihilation into neutrino-antineutrino pairs would reduce the DM
energy density to a very tiny value today. An interesting
alternative possibility is that the relic DM abundance is the
outcome of a particle-antiparticle asymmetry produced at higher
temperatures in the DM sector, very much akin to the mechanism by
which the baryon number is produced in the early Universe. Indeed,
in this case one might also account for the fact that intriguingly
the two energy density parameters for baryons and DM, $\Omega_b$
and $\Omega_{DM}$ only differ by a factor five today, yet their
production mechanism is usually considered to be quite distinct.

To conclude, it is interesting to compare the bounds on $Q_2$ and
$Q_0$ with those which can be obtained from the propagation of
astrophysical neutrinos. The most important constraint is provided
by observation of neutrinos from SN1987A. These neutrinos have
energies of order 10 MeV.  The thickness of the dark matter layer
that they propagate through is approximately $\int \rho(l) dl$,
the integral of the DM density along the line of sight $l$ to the
LMC. Approximating the DM density $\rho(l) \sim \rho_0
(l/l_0)^{-2}$, where $\rho_0\simeq0.4$ GeV~cm$^{-3}$ is the local
density and $l_0\simeq8$ kpc our distance from the Galactic
center, we find a dark matter thickness $\sim10^{25}$
MeV~cm$^{-2}$.  Given the agreement between the predicted and
observed neutrino flux and energy spectrum, one infers that
neutrinos from SN1987A were not significantly absorbed by dark
matter along the line of sight, from which one gets $Q_{2} \le
10^{-45}\, \mbox{cm}^2~\mbox{MeV}^{-1}$ for both $m_{DM}$ and
$m_F$ in the MeV range, and $Q_{0} \le \,10^{-25}\,
\mbox{cm}^{2}~\mbox{MeV}^{-1}$. Note that the bound on $Q_2$ is
stronger than what is obtained using LSS data, while for $Q_0$ the
stronger bound is still provided by LSS.\\

\noindent {\bf ACKNOWLEDGMENTS}\\

I'm pleased to thank the Organizers and Conveners of the NOW 2006
Workshop for their successful efforts in creating a fruitful and
relaxed atmosphere during the Conference. I also thank my
collaborators who enjoyed with me the study of this subject
discussed in more details in \cite{prd}.

\end{document}